\documentclass[preprint, 12pt, authoryear, nonatbib]{elsarticle}

\usepackage[margin=1in]{geometry} 

\usepackage[utf8]{inputenc}
\usepackage[T1]{fontenc}
\usepackage[english]{babel}
\usepackage{csquotes}
\usepackage[style=numeric-comp, sorting=none, natbib=true, backend=biber]{biblatex} 
\usepackage{amsmath}          
\usepackage{amssymb}          
\usepackage{amsthm}           
\usepackage{mathtools}        
\usepackage{bm}               
\usepackage{siunitx}          
\usepackage{tensor}           

\usepackage{graphicx}         
  \graphicspath{{figures/}}   
\usepackage{float}            
\usepackage{placeins}         
\usepackage{subcaption}       
\usepackage{wrapfig}          

\usepackage{booktabs}         
\usepackage{array}            
\usepackage{multirow}         
\usepackage{multicol}         
\usepackage{tabularx}         
\usepackage{longtable}        
\usepackage{makecell}         

\usepackage[dvipsnames, table, xcdraw]{xcolor}
  \definecolor{ader-blue}{RGB}{31,  56, 100}
  \definecolor{ader-green}{RGB}{39, 80,  17}
  \definecolor{ader-red}{RGB}{163, 45,  45}

\usepackage[hidelinks]{hyperref}
\usepackage{cleveref}         
  \crefformat{equation}{(#2#1#3)}   
\usepackage{url}              

\usepackage[ruled, vlined, linesnumbered]{algorithm2e}
\usepackage{listings}
\usepackage{tikz}
  \usetikzlibrary{arrows.meta, positioning, shapes.geometric,
                  decorations.pathreplacing, calc, patterns}
\usepackage{pgfplots}
  \pgfplotsset{compat=1.18}
  \usepgfplotslibrary{groupplots, colorbrewer}

\usepackage{lineno}           
\usepackage[colorinlistoftodos,
            prependcaption,
            textsize=tiny]{todonotes}
\usepackage{lipsum}           

\usepackage[nopatch=footnote]{microtype}
\usepackage{comment}          
\usepackage{appendix}         
\usepackage{setspace}         
\newcommand{\aderdg}{\textsc{ader-dg}}
\newcommand{\exahype}{\textsc{ExaHyPE}}

\newcommand{\vect}[1]{\boldsymbol{#1}}         
\newcommand{\mat}[1]{\mathbf{#1}}              

\journal{Astronomy and Computing}

\begin{document}

\begin{frontmatter}

\title{High-Order \textsc{ader-dg} Hydrodynamics with \textsc{ExaHyPE}:
       Implementation, Validation, and Astrophysical Benchmarking}

\author[unal,bosque]{Andr\'{e}s Mauricio Su\'{a}rez Mantilla%
                      \corref{cor1}}
\ead{anmsuarezma@unal.edu.co}
\cortext[cor1]{Corresponding author}

\author[oan]{Leonardo Casta\~{n}eda Colorado}
\ead{lcastanedac@unal.edu.co}

\affiliation[unal]{
  organization = {Universidad Nacional de Colombia},
}
\affiliation[bosque]{
  organization = {Universidad El Bosque},
}
\affiliation[oan]{
  organization = {Observatorio Astron\'{o}mico Nacional,
                  Universidad Nacional de Colombia},
}

\begin{abstract}
We describe a high-order \aderdg{} solver for the compressible Euler
equations within the \exahype{} framework. The implementation combines
a high-order \aderdg{} polynomial representation, a local space--time DG
predictor, adaptive mesh refinement, and an \textit{a posteriori} subcell
finite-volume limiter. We test the code on a deliberately mixed set of
one- and two-dimensional problems: a strong-shock Sod-type problem, the
Shu--Osher shock--entropy interaction, the Woodward--Colella blast wave,
a contact-driven vortex sheet, and a shock--interface interaction. The
one-dimensional cases recover the expected Euler wave patterns and show
clear order-dependent gains in smooth and oscillatory regions. The
two-dimensional cases probe a different part of the method, namely
contact preservation, shear-driven roll-up, baroclinic vorticity
deposition, and Richtmyer--Meshkov-type growth. In these tests the
high-order update gives the expected resolution away from discontinuities,
whereas the subcell limiter keeps the calculation stable near shocks and
steep interfaces. The resulting code provides a reproducible \exahype{}
implementation for idealised inviscid, non-relativistic flows in which
shocks, contacts, and multidimensional interfaces are the dominant
features. Codes and datasets are publicly available.

\end{abstract}


\end{frontmatter}


\section{Introduction}
\label{sec:introduction}

Compressible-flow simulations are difficult because smooth waves,
shocks, and contact discontinuities can appear in the same calculation.
This is especially relevant in gas dynamics and astrophysical
hydrodynamics, where strong compressions and interacting interfaces are
part of the flow rather than exceptional events. Mathematically, these
problems are governed by hyperbolic systems of conservation laws, whose
solutions may develop discontinuities even from smooth initial data
\citep{LeVeque1992, Toro2009}. The numerical difficulty is therefore
not only to obtain high formal order in smooth regions. The method must
also remain conservative and non-oscillatory when shocks or contacts are
formed. This tension has been present since the work of
\citet{Godunov1959}: low-order schemes are robust but often too
diffusive, while high-order schemes require additional mechanisms to
control oscillations without erasing the structures one wants to
resolve.

High-order shock-capturing methods address this balance directly. The
ADER methodology of \citet{ToroTitarev2002} provided a one-step framework
for high-order accuracy in both space and time through the solution of
Generalised Riemann Problems. Later extensions connected this idea with
multidimensional nonlinear systems and with the Discontinuous Galerkin
framework
\citep{TitarevToro2005, CockburnShu1998, DumbserKaser2007, Dumbser2008}.
For the present work, the important point is the modern ADER-DG form: a
high-order DG polynomial represents the cell-local solution, a local
space-time DG predictor evolves that polynomial inside each cell, and a
conservative corrector communicates neighbouring cells through numerical
fluxes.

The framework has also become more robust in shock-dominated regimes.
The \textit{a posteriori} subcell finite-volume limiter of
\citet{Dumbser2014} allows the high-order DG candidate to be tested
after it is computed and replaced only where it becomes inadmissible.
Space-time adaptive formulations with dynamic adaptive mesh refinement
\citep{Zanotti2015a} add a second practical ingredient: resolution can
be concentrated near shocks, contacts, and emerging small-scale
structures. These developments have supported applications in
relativistic magnetohydrodynamics, compressible Navier--Stokes and
resistive magnetohydrodynamics, general-relativistic continuum
mechanics, and the Einstein--Euler system
\citep{Zanotti2015b, Fambri2017, Peshkov2019, Dumbser2024}. In this
context, the \exahype{} engine provides a computational environment for
high-order hyperbolic solvers with parallelism and dynamic adaptivity
\citep{Reinarz2020}. Many published \exahype{} applications have
focused on geophysical wave propagation and relativistic astrophysical
fluid dynamics \citep{Duru2022, Reinarz2020}.

The non-relativistic Euler equations have received less attention in
this particular setting. This absence matters for two reasons. First,
the Euler system remains the natural starting point for compressible
hydrodynamics and for controlled tests of shock-capturing methods.
Second, the numerical ingredients that make \aderdg{} attractive in more
complex models---high-order polynomial representation, local space-time
prediction, adaptive refinement, and subcell limiting---can be examined
more cleanly before additional physics is introduced. A systematic
Euler implementation in \exahype{} is therefore not just a simplified
case; it is a necessary step for understanding how these ingredients
behave together. To the best of our knowledge, the published literature
does not yet provide a reproducible implementation in this environment
that combines a high-order \aderdg{} polynomial representation, a
space-time DG predictor, adaptive mesh refinement, and an
\textit{a posteriori} subcell limiter with a validation suite for one- and
two-dimensional Euler benchmarks.

In this work, we present such a solver for the compressible Euler
equations and assess its behaviour through benchmark problems of
increasing complexity. In one spatial dimension, we consider the Sod
shock tube \citep{Sod1978}, the Shu--Osher entropy-wave interaction
\citep{ShuOsher1989}, and the Woodward--Colella blast-wave test
\citep{WoodwardColella1984}. In two dimensions, we study interacting
contact discontinuities and a shock--interface configuration that
produces multiscale structures associated with Richtmyer--Meshkov and
Kelvin--Helmholtz dynamics. With this set of tests, we do not only
verify the implementation. We also examine how the polynomial representation,
space-time prediction, limiting, and adaptivity behave when smooth
waves, sharp discontinuities, and multidimensional effects coexist.
Several of these configurations are also relevant to astrophysical
hydrodynamics, where non-relativistic shock interactions remain an
important modelling regime.

All simulation codes and datasets produced in this study are made
publicly available to support reproducibility and future extensions.
The paper is organised as follows. Section~\ref{sec:governing_equations}
introduces the Euler equations and their characteristic structure.
Section~\ref{sec:numerical_method} describes the \aderdg{} method and
its limiting strategy, and Section~\ref{sec:implementation} summarises
the implementation in \exahype{}. Sections~\ref{sec:results_1d}
and~\ref{sec:results_2d} present the one- and two-dimensional results.
Section~\ref{sec:discussion} discusses the main findings, and
Section~\ref{sec:conclusions} closes the paper.

\section{Governing Equations}
\label{sec:governing_equations}

\subsection{The Compressible Euler Equations}
\label{sec:euler_equations}

We model the flow with the inviscid compressible Euler equations in
conservative form. This is the form needed once shocks appear, because
the solution must then be interpreted weakly and the numerical method
must preserve conservation across discontinuities. The equations read
\citep{LeVeque1992, Toro2009}
\begin{equation}
  \frac{\partial \vect{U}}{\partial t}
  + \nabla \cdot \mat{F}(\vect{U}) = \vect{0},
  \label{eq:euler_conservative}
\end{equation}
where $\vect{U} \in \mathbb{R}^{m}$ is the vector of conserved
variables and $\mat{F}:\mathbb{R}^m\to\mathbb{R}^{m\times d}$ is the
flux tensor, with $m=d+2$ in $d$ spatial dimensions. In one dimension,
\begin{equation}
  \vect{U} =
  \begin{pmatrix} \rho \\ \rho u \\ E \end{pmatrix},
  \quad
  \mat{F}(\vect{U}) =
  \begin{pmatrix} \rho u \\ \rho u^2 + p \\ (E+p)u \end{pmatrix},
  \label{eq:UF_1d}
\end{equation}
while in two dimensions we use
\begin{equation}
  \vect{U} =
  \begin{pmatrix} \rho \\ \rho u \\ \rho v \\ E \end{pmatrix},
  \quad
  \mat{F}_1 =
  \begin{pmatrix} \rho u \\ \rho u^2 + p \\ \rho uv \\ (E+p)u \end{pmatrix},
  \quad
  \mat{F}_2 =
  \begin{pmatrix} \rho v \\ \rho uv \\ \rho v^2 + p \\ (E+p)v \end{pmatrix}.
  \label{eq:UF_2d}
\end{equation}
Here $\rho>0$ is the density, $(u,v)$ are the velocity components,
$E$ is the total energy per unit volume, and $p$ is the thermal
pressure. The physically admissible states are those for which density
and pressure remain positive,
\begin{equation}
  \mathcal{U} = \bigl\{ \vect{U} \in \mathbb{R}^{m} :
  \rho > 0,\; p(\vect{U}) > 0 \bigr\}.
  \label{eq:state_space}
\end{equation}
This set is important later because the limiter is designed precisely
to reject candidate updates that leave $\mathcal{U}$.

\subsection{Equation of State and Admissible Thermodynamic States}
\label{sec:eos}

We close the system with the ideal-gas equation of state
\citep{LandauLifshitz1987, Toro2009},
\begin{equation}
  p = (\gamma - 1)\,\rho\varepsilon,
  \label{eq:eos}
\end{equation}
where $\varepsilon$ is the specific internal energy and
$\gamma=c_p/c_v>1$ is the adiabatic index. The total energy density is
\begin{equation}
  E = \rho\varepsilon + \tfrac{1}{2}\rho |\vect{u}|^2,
  \label{eq:total_energy}
\end{equation}
so the pressure recovered from conserved variables is
\begin{equation}
  p(\vect{U}) = (\gamma - 1)
  \Bigl(E - \tfrac{1}{2}\rho |\vect{u}|^2\Bigr).
  \label{eq:pressure_recovery}
\end{equation}
This is the quantity checked in the positivity criterion of the
limiter. The corresponding sound speed is
\begin{equation}
  a = \sqrt{\frac{\gamma p}{\rho}},
  \label{eq:sound_speed}
\end{equation}
which sets the acoustic propagation scale. Throughout this work we use
$\gamma=7/5$, except in benchmark problems where another value is
specified.

\subsection{Hyperbolicity and Characteristic Wave Structure}
\label{sec:hyperbolicity}

For the characteristic analysis, the one-dimensional Euler equations
are written in quasi-linear form as
\begin{equation}
  \frac{\partial \vect{U}}{\partial t}
  + \mat{A}(\vect{U})\,\frac{\partial \vect{U}}{\partial x}
  = \vect{0},
  \qquad
  \mat{A}(\vect{U}) = \frac{\partial \mat{F}}{\partial \vect{U}}.
  \label{eq:quasilinear}
\end{equation}
In the admissible state space, this system is strictly hyperbolic
\citep{LeVeque1992, Toro2009}. Its eigenvalues are
\begin{equation}
  \lambda_1 = u-a, \qquad
  \lambda_2 = u, \qquad
  \lambda_3 = u+a,
  \label{eq:eigenvalues}
\end{equation}
with $\lambda_1<\lambda_2<\lambda_3$ whenever $a>0$. The corresponding
right eigenvectors form
\begin{equation}
  \mat{R}(\vect{U})
  = \bigl[\,\vect{r}_1 \;\big|\; \vect{r}_2 \;\big|\; \vect{r}_3\,\bigr]
  =
  \begin{pmatrix}
    1      & 1               & 1      \\
    u - a  & u               & u + a  \\
    H - ua & \tfrac{1}{2}u^2 & H + ua
  \end{pmatrix},
  \label{eq:right_eigenvectors}
\end{equation}
where $H=(E+p)/\rho$ is the specific total enthalpy. With
$\mat{L}=\mat{R}^{-1}$, the flux Jacobian is diagonalised as
\begin{equation}
  \mat{A}(\vect{U})
  = \mat{R}(\vect{U})\,\mat{\Lambda}(\vect{U})\,\mat{L}(\vect{U}),
  \qquad
  \mat{\Lambda}=\mathrm{diag}(\lambda_1,\lambda_2,\lambda_3).
  \label{eq:diagonalisation}
\end{equation}
This characteristic basis also fixes the local wave decomposition used by
the numerical fluxes and admissibility checks. Locally, the characteristic
variables
\begin{equation}
  \frac{\partial \vect{W}}{\partial t}
  + \mat{\Lambda}\,\frac{\partial \vect{W}}{\partial x}
  \approx \vect{0},
  \qquad
  \vect{W}=\mat{L}\,\vect{U},
  \label{eq:characteristic_system}
\end{equation}
separate the solution into wave families that propagate with speeds
$\lambda_k$.

This decomposition also tells us what the numerical method must
capture. The acoustic fields associated with $\lambda_1$ and
$\lambda_3$ are genuinely nonlinear, whereas the middle field
associated with $\lambda_2$ is linearly degenerate \citep{Lax1957}.
The outer families generate shocks or rarefactions; the middle family
corresponds to a contact discontinuity. In physical terms, acoustic
waves carry pressure information, while the contact wave transports
material and thermodynamic jumps at the fluid speed. This distinction
is the reason why the benchmark suite must include both shocks and
contact-dominated configurations.

Once shocks form, the solution is no longer classical. For the present
purposes, two facts are enough. Conservative formulations remain
meaningful after discontinuities appear, and physically relevant shocks
must satisfy an entropy admissibility criterion. For a $k$-shock of
speed $\sigma$, the Lax condition is
\begin{equation}
  \lambda_k(\vect{U}_R) < \sigma < \lambda_k(\vect{U}_L),
  \label{eq:lax_entropy}
\end{equation}
which states that characteristics of the $k$-th family enter the shock
from both sides \citep{Lax1957, LeVeque1992}. This is the continuous
background behind the shock-capturing and admissibility checks used
later.

In two dimensions, the same wave structure is recovered along a chosen
normal direction. The quasi-linear form is
\begin{equation}
  \frac{\partial \vect{U}}{\partial t}
  + \mat{A}_x(\vect{U})\,\frac{\partial \vect{U}}{\partial x}
  + \mat{A}_y(\vect{U})\,\frac{\partial \vect{U}}{\partial y}
  = \vect{0},
  \label{eq:quasilinear_2d}
\end{equation}
with $\mat{A}_x=\partial\mat{F}_1/\partial\vect{U}$ and
$\mat{A}_y=\partial\mat{F}_2/\partial\vect{U}$. For a unit normal
$\hat{\vect{n}}=(n_x,n_y)^\top$, the projected Jacobian
$\mat{A}_{\hat{n}}=n_x\mat{A}_x+n_y\mat{A}_y$ has eigenvalues
\begin{equation}
  \lambda_{1,\hat{n}} = \vect{u}\cdot\hat{\vect{n}} - a,
  \qquad
  \lambda_{2,3,\hat{n}} = \vect{u}\cdot\hat{\vect{n}},
  \qquad
  \lambda_{4,\hat{n}} = \vect{u}\cdot\hat{\vect{n}} + a.
  \label{eq:eigenvalues_2d}
\end{equation}
The system is therefore hyperbolic in every spatial direction. The
middle eigenvalue is double because tangential momentum adds one
material degree of freedom, but the eigenspace remains complete. This
normal-direction structure is the one used in interface fluxes and in
the CFL estimate of the numerical method.

\subsection{The Riemann Problem and its Generalisation}
\label{sec:riemann_problem}

The numerical fluxes are built around the same local wave structure.
At a cell interface, the basic model is the Riemann problem,
\begin{equation}
  \vect{U}(\vect{x},0) =
  \begin{cases}
    \vect{U}_L & \text{if } x<0, \\
    \vect{U}_R & \text{if } x>0,
  \end{cases}
  \label{eq:riemann_ic}
\end{equation}
where two constant states meet at a single discontinuity. In one
dimension its solution consists of two acoustic waves and one contact
wave: the acoustic waves may be shocks or rarefactions, while the
contact carries density and entropy jumps at continuous pressure and
normal velocity \citep{Toro2009}. This is the local pattern behind the
shock-tube and interaction problems used later.

This local problem also fixes the numerical point of view. Godunov's
method obtains a conservative upwind flux from it \citep{Godunov1959}.
ADER keeps the same interface-centred idea, but replaces piecewise
constant states with local polynomial data and evolves them in time.
This Generalised Riemann Problem \citep{BenArtzi1984} is what lets ADER
schemes reach high order in space and time without multi-stage time
integration \citep{ToroTitarev2002}. In the next section, this local
view is combined with the high-order \aderdg{} polynomial
representation, the space-time predictor, and the conservative
corrector.

\section{Numerical Method}
\label{sec:numerical_method}

The numerical tests require a method that remains high order in smooth
regions while staying robust at shocks and contact discontinuities. We
therefore use the limiting \aderdg{} formulation available in
\exahype{}: a high-order DG polynomial represents the solution inside
each cell, a local space--time predictor evolves that polynomial over one
time step, and a conservative corrector couples neighbouring cells through
numerical fluxes. If the high-order candidate fails admissibility checks,
an \textit{a posteriori} subcell finite-volume limiter recomputes only
the troubled cells.

\subsection{High-Order DG Polynomial Representation}
\label{sec:dg_polynomial}

At time $t^n$, the numerical solution in each cell $C_i$ is represented
by a polynomial $u_h^n$ of degree $N$. In reference coordinates
$\zeta\in[0,1]$, we write
\begin{equation}
  u_h^n(\zeta) = \sum_{p=0}^{N} \hat{u}_{p}^{\,n}\,\phi_p(\zeta),
  \label{eq:dg_polynomial}
\end{equation}
where $\{\phi_p\}_{p=0}^{N}$ are local nodal basis functions and
$\hat{u}_{p}^{\,n}$ are the cell-local degrees of freedom. The same
polynomial representation is used for each conserved variable of the
Euler system. Increasing $N$ raises the formal order of the smooth-cell
update, while the limiting strategy described below protects the
calculation when the candidate polynomial becomes inadmissible near
discontinuities.

\subsection{Local Space-Time DG Predictor}
\label{sec:predictor}

The cell-local DG polynomial is then evolved locally over
$C_i\times[t^n,t^{n+1}]$. The result is a space-time polynomial
$q_h(\zeta,\tau)$ computed without neighbour communication. This local
predictor replaces the analytic Cauchy--Kowalewski expansion of early
ADER schemes \citep{ToroTitarev2002} and is better suited to nonlinear
systems \citep{Dumbser2008}.

With reference coordinates
\begin{equation}
  x = x_{i-1/2} + \zeta\,\Delta x_i,
  \qquad
  t = t^n + \tau\,\Delta t,
  \qquad
  (\zeta,\tau) \in [0,1]^2,
  \label{eq:ref_coords}
\end{equation}
the predictor is expanded as
\begin{equation}
  q_h(\zeta,\tau) = \theta_P(\zeta,\tau)\,\hat{q}_P,
  \qquad
  \theta_P(\zeta,\tau) = \psi_p(\zeta)\psi_q(\tau),
  \label{eq:predictor_ansatz}
\end{equation}
with summation over repeated indices. Projecting
$\partial_\tau q_h+\partial_\zeta f^*(q_h)=0$, where
$f^*=(\Delta t/\Delta x_i)f$, and integrating the time derivative by
parts gives
\begin{equation}
\begin{aligned}
  &\int_0^1 \theta_Q(\zeta,1)q_h(\zeta,1)\,\mathrm{d}\zeta
  - \int_0^1\!\!\int_0^1
    \frac{\partial\theta_Q}{\partial\tau}q_h\,
    \mathrm{d}\tau\,\mathrm{d}\zeta \\
  &\quad + \int_0^1\!\!\int_0^1
    \theta_Q\frac{\partial f^*(q_h)}{\partial\zeta}\,
    \mathrm{d}\zeta\,\mathrm{d}\tau  \\
  &= \int_0^1 \theta_Q(\zeta,0)u_h^n(\zeta)\,
    \mathrm{d}\zeta,
  \label{eq:predictor_weak}
\end{aligned}
\end{equation}
where $u_h^n$ is the DG polynomial at $\tau=0$. The required
matrices are
\begin{equation}
\begin{aligned}
  K^1_{QP}
  &= \int_0^1 \theta_P(\zeta,1)\theta_Q(\zeta,1)\,
     \mathrm{d}\zeta, \\
  K^\tau_{QP}
  &= \int_0^1\!\!\int_0^1
     \frac{\partial\theta_Q}{\partial\tau}\theta_P\,
     \mathrm{d}\tau\,\mathrm{d}\zeta, \\
  K^\zeta_{QP}
  &= \int_0^1\!\!\int_0^1
     \theta_Q\frac{\partial\theta_P}{\partial\zeta}\,
     \mathrm{d}\zeta\,\mathrm{d}\tau.
  \label{eq:stiffness_matrices}
\end{aligned}
\end{equation}
The predictor degrees of freedom satisfy
\begin{equation}
  \bigl(K^1_{QP}-K^\tau_{QP}\bigr)\hat{q}^{\,n+1}_P
  = -K^\zeta_{QP}\hat{f}^{\,*}_P(\hat{q}_P) + F^0_{QP},
  \label{eq:predictor_system}
\end{equation}
with $F^0_{QP}=\int_0^1\theta_Q(\zeta,0)u_h^n(\zeta)\,\mathrm{d}\zeta$.
It is solved by Picard iteration,
\begin{equation}
\begin{aligned}
  \hat{q}^{(l+1)}_P
  &= \bigl(K^1_{QP}-K^\tau_{QP}\bigr)^{-1} \\
  &\quad \times
     \left[-K^\zeta_{QP}\hat{f}^{\,*}_P(\hat{q}^{(l)}_P)
     + F^0_{QP}\right].
\end{aligned}
\end{equation}
For the CFL-limited steps used here, only a few iterations are needed
\citep{Dumbser2008}. The predictor then provides the time-dependent
interface traces at $\zeta=0$ and $\zeta=1$ for the conservative update.

\subsection{ADER-DG Corrector Step}
\label{sec:corrector}

The corrector is where neighbouring cells communicate. From the left
and right predictor traces, $q_h^-$ and $q_h^+$, the cell average is
advanced in one conservative step:
\begin{equation}
\begin{aligned}
  \hat{\vect{U}}^{n+1}_i
  &= \vect{U}^n_i
  - \frac{\Delta t}{\Delta x_i}\int_0^1
    \Bigl[
    \mat{F}_{\mathrm{RS}}\!\left(q_h^-(x_{i+1/2},\tau),
                                 q_h^+(x_{i+1/2},\tau)\right) \\
  &\qquad\qquad\qquad
    - \mat{F}_{\mathrm{RS}}\!\left(q_h^-(x_{i-1/2},\tau),
                                   q_h^+(x_{i-1/2},\tau)\right)
    \Bigr] \mathrm{d}\tau .
  \label{eq:ader_update}
\end{aligned}
\end{equation}
Here $\mat{F}_{\mathrm{RS}}$ is the Rusanov, or local
Lax--Friedrichs, flux, and the time integral is evaluated with
Gauss--Legendre quadrature. The result $\hat{\vect{U}}^{n+1}_i$ is a
high-order candidate: it is accepted in smooth regions but checked
before being committed.

The time step follows the DG CFL restriction
\begin{equation}
  \Delta t \leq \mathrm{CFL}\,
  \frac{\Delta x_{\min}}{(2N+1)\,|\lambda|_{\max}},
  \label{eq:cfl}
\end{equation}
where $|\lambda|_{\max}=|\vect{u}\cdot\hat{\vect{n}}|+a$,
$\Delta x_{\min}$ is the smallest local cell size, and $(2N+1)^{-1}$
accounts for the polynomial resolution inside a cell. Unless a
benchmark states otherwise, $\mathrm{CFL}=0.9$.

\subsection{A Posteriori Subcell Finite-Volume Limiter}
\label{sec:limiter}

\begin{figure}[H]
  \centering
  \includegraphics[width=0.95\textwidth]{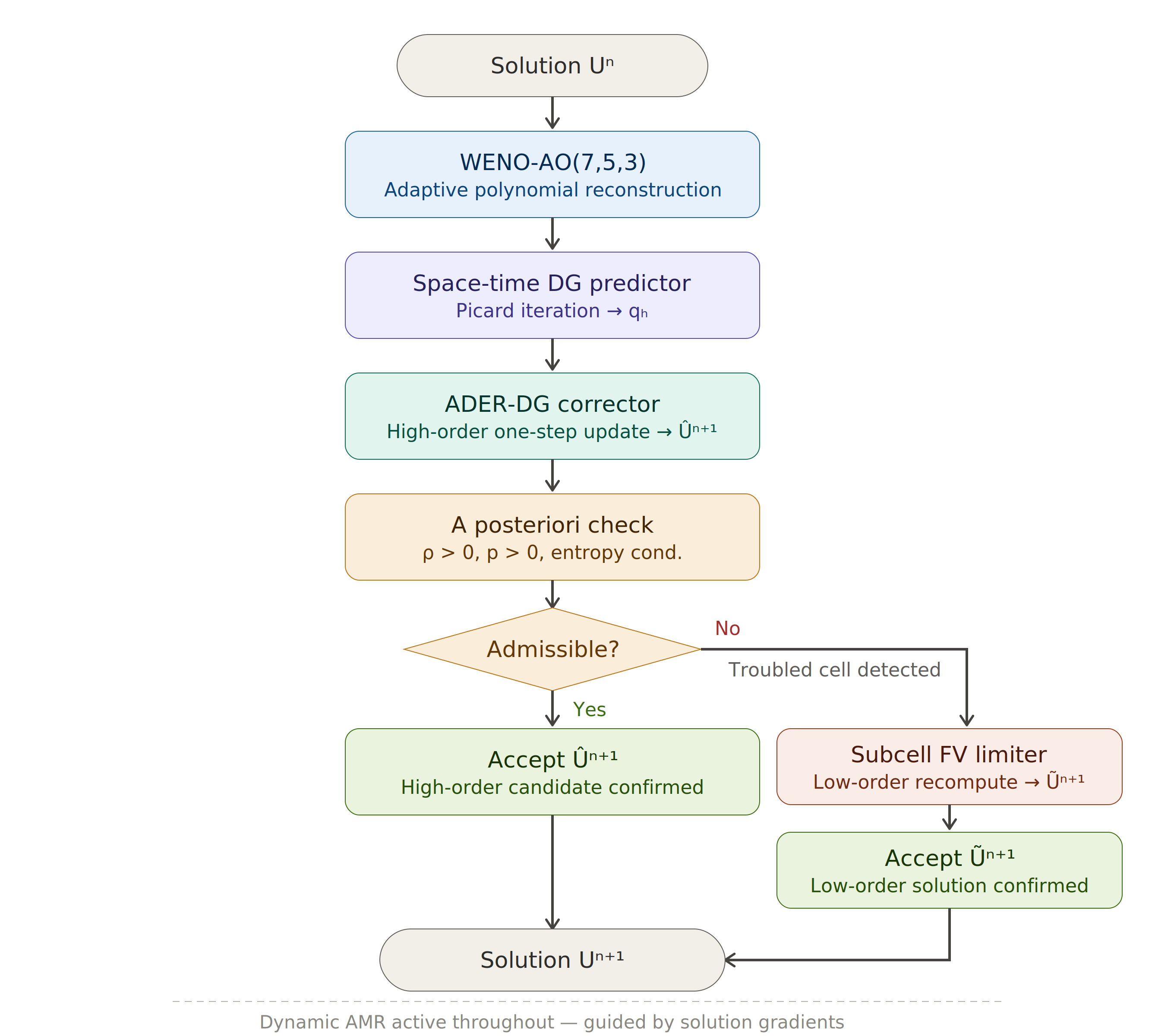}
  \caption{Workflow of the \aderdg{} update with \textit{a posteriori}
  subcell finite-volume limiting. The scheme first constructs a
  high-order candidate, checks its admissibility, and recomputes only
  troubled cells on a finite-volume subgrid.}
  \label{fig:aderdg_flowchart}
\end{figure}

Near strong discontinuities, the high-order candidate may contain
negative density, negative pressure, or non-finite values. We use the
\textit{a posteriori} strategy of \citet{Dumbser2014}: compute the
high-order candidate, test it, and recompute only the cells that fail.
This keeps the limiter local and preserves the unlimited scheme where
it is already admissible. The workflow is shown in
Figure~\ref{fig:aderdg_flowchart}.

\FloatBarrier

Each candidate $\hat{\vect{U}}^{n+1}_i$ is checked against three
criteria:
\begin{enumerate}
  \item Physical positivity: $\rho(\hat{\vect{U}}^{n+1}_i)>0$ and
        $p(\hat{\vect{U}}^{n+1}_i)>0$, with pressure recovered from
        \eqref{eq:pressure_recovery};
  \item Entropy admissibility: the local entropy measure
        $s=p/\rho^\gamma$ is compared with neighbourhood values at
        time $t^n$;
  \item Numerical regularity: the candidate must not contain
        not-a-number or infinite values.
\end{enumerate}

A cell that passes keeps the high-order candidate. A cell that fails is
marked as troubled and recomputed on $N_s=2N+1$ equispaced subcells
\citep{Dumbser2014, Zanotti2015a}. The previous cell average is
projected to subcell averages, evolved with a second-order TVD
finite-volume method using the Rusanov flux and a minmod slope limiter,
and projected back to obtain $\tilde{\vect{U}}^{n+1}_i$. This limited
state replaces the candidate only in the troubled cell. The choice
$N_s=2N+1$ lets the subcell update use the same time step as the DG
scheme.

The limiter is deliberately conservative: it leaves admissible
high-order updates untouched, but provides a robust fallback near
shocks, contacts, and strong compressions. The troubled-cell set also
acts as a diagnostic for regions requiring additional numerical control.

\subsection{Adaptive Mesh Refinement}
\label{sec:amr}

For adaptive runs, we use the dynamic mesh refinement available in
\exahype{} \citep{Reinarz2020}, following the space-time adaptive
formulation of \citet{Zanotti2015a}. Refinement is driven by local
variation in the DG solution, and coarsening is applied
where that variation falls below the prescribed tolerance. The
\aderdg{} update and subcell limiter are applied independently in each
AMR patch, with the local time step constrained by \eqref{eq:cfl}.

Fluxes across coarse--fine interfaces are corrected conservatively, so
refinement does not compromise global conservation. AMR therefore
concentrates resolution near shocks and contacts while keeping smooth
regions inexpensive.

\section{Implementation in \exahype{}}
\label{sec:implementation}

The method is implemented in \exahype{}, an open-source C++ framework
for parallel, dynamically adaptive solvers for hyperbolic PDEs
\citep{Reinarz2020}. The framework provides pure \aderdg{}, pure
finite-volume, and \texttt{LimitingADERDG} solvers. We use
\texttt{LimitingADERDG} because it matches the \textit{a posteriori}
subcell strategy described in Section~\ref{sec:limiter}. Although
OpenMP and MPI are available, the benchmarks reported here run on a
single shared-memory node.

\subsection{Solver Configuration}
\label{sec:solver_config}

The Euler solver is specified through six C++ kernel functions. The
routine \texttt{flux} evaluates the flux tensor $\mat{F}(\vect{U})$ in
equations~\eqref{eq:UF_1d}--\eqref{eq:UF_2d}, with pressure recovered
from~\eqref{eq:pressure_recovery}. The routine \texttt{eigenvalues}
returns $|\lambda|_{\max}=|\vect{u}\cdot\hat{\vect{n}}|+a$ in each
coordinate direction; this is the speed used in the CFL
condition~\eqref{eq:cfl}. The routine \texttt{eigenvectors} provides
$\mat{R}(\vect{U})$ and $\mat{L}(\vect{U})$ from
\eqref{eq:right_eigenvectors} for the characteristic decomposition used
by the Euler kernels. Finally,
\texttt{boundaryValues} sets the boundary conditions described in
Section~\ref{sec:bcs}. Together, these routines define the physical
part of the solver.

The limiter is controlled by the two remaining routines.
\texttt{mapDiscrete\discretionary{}{}{}Maximum\discretionary{}{}{}Principle\discretionary{}{}{}Observables} declares
$\rho$ and $p$ as the monitored quantities. The routine
\texttt{isPhysically\discretionary{}{}{}Admissible} implements the checks of
Section~\ref{sec:limiter}: it rejects states with $\rho\leq0$,
$p\leq0$, a non-finite variable, or an entropy measure
$s=p/\rho^\gamma$ below the neighbourhood minimum at $t^n$. A
\texttt{false} return triggers subcell recomputation for that cell.

The numerical parameters common to all simulations are listed in
Table~\ref{tab:parameters}. Polynomial degree $N$ and AMR levels are
benchmark-specific and are stated in Sections~\ref{sec:results_1d}
and~\ref{sec:results_2d}.

\begin{table}[htbp]
  \centering
  \caption{Numerical parameters used unless stated otherwise.
           Polynomial degree $N$ and AMR levels are benchmark-specific;
           see Sections~\ref{sec:results_1d} and~\ref{sec:results_2d}.}
  \label{tab:parameters}
  \begin{tabular}{llll}
    \toprule
    Parameter & Symbol & Value & Reference \\
    \midrule
    CFL number
      & $\mathrm{CFL}$
      & $0.9$
      & Eq.~\eqref{eq:cfl} \\
    Subcell count
      & $N_s$
      & $2N+1$
      & Section~\ref{sec:limiter} \\
    Temporal quadrature
      & Gauss--Legendre
      & $N+1$ points
      & Section~\ref{sec:corrector} \\
    Spatial quadrature
      & Gauss--Legendre
      & $N+1$ points
      & Section~\ref{sec:predictor} \\
    Interface flux
      & Rusanov (LLF)
      & $\alpha = |\lambda|_{\max}$
      & Section~\ref{sec:corrector} \\
    Subcell slope limiter
      & minmod
      & \textemdash{}
      & Section~\ref{sec:limiter} \\
    Adiabatic index
      & $\gamma$
      & $7/5$
      & Eq.~\eqref{eq:eos} \\
    \bottomrule
  \end{tabular}
\end{table}

\subsection{Initial and Boundary Conditions}
\label{sec:bcs}

Initial conditions are set through \texttt{adjustPointSolution}, which
is called at $t=0$ for each mesh degree of freedom. The benchmark
states are given in Sections~\ref{sec:results_1d}
and~\ref{sec:results_2d}.

Boundary conditions use one of two strategies. One-dimensional tests
use transmissive outflow boundaries at both ends, implemented by
copying the adjacent interior state into ghost cells so that waves can
leave the domain without reflection. Two-dimensional tests use periodic
conditions transverse to the shock propagation direction and
transmissive conditions along the propagation direction. Both choices
are implemented through \texttt{boundaryValues} and selected in the
\exahype{} configuration file, keeping benchmark-specific data separate
from the solver kernels.

\section{One-Dimensional Benchmark Tests}
\label{sec:results_1d}

The sequence of tests is intentional. We begin with the Sod
shock tube, where an exact solution exercises all three wave
families. We then move to Shu--Osher, where a shock interacts
with fine-scale entropy structure, and finally to the
Woodward--Colella blast wave, where strong shocks reflect and
collide inside a confined domain.
All simulations use the parameters of Table~\ref{tab:parameters}.
Benchmark-specific values are stated below.

\subsection{Sod Shock Tube}
\label{sec:sod}

The Sod shock tube is a Riemann problem with a known exact
solution. Two uniform states meet at a diaphragm placed at
$x = 0.5$. When the diaphragm is removed, the solution
develops a left-going rarefaction fan, a contact discontinuity,
and a right-going shock — one representative of each wave
family from Section~\ref{sec:hyperbolicity}. The initial
conditions are in Table~\ref{tab:sod_ic}.

\begin{table}[htbp]
  \centering
  \caption{Initial conditions for the Sod shock tube.
           Domain $[0,1]$, final time $t = 0.012$.}
  \label{tab:sod_ic}
  \begin{tabular}{lccc}
    \toprule
    State & $\rho$ & $u$ & $p$ \\
    \midrule
    Left  & $1.0$ & $0.0$ & $1000.0$ \\
    Right & $1.0$ & $0.0$ & $0.01$   \\
    \bottomrule
  \end{tabular}
\end{table}

We discretise the domain with maximum mesh size $\Delta x = 0.1$
and two AMR levels, giving an effective resolution of
$\Delta x_{\min} = 0.025$. We run four schemes at polynomial
orders $N = 2, 4, 6, 8$, corresponding to third- through
ninth-order accuracy. Transmissive outflow conditions are
imposed at both ends.

This is a strong-shock Sod-type variant rather than the
classical textbook Sod test. The pressure ratio
$p_L/p_R = 10^5$ is chosen deliberately to stress the
high-order update and the subcell limiter under a compressive
contrast closer to idealised blast-wave and supernova-remnant
conditions.

\begin{figure}[H]
  \centering
  \includegraphics[width=0.64\textwidth]{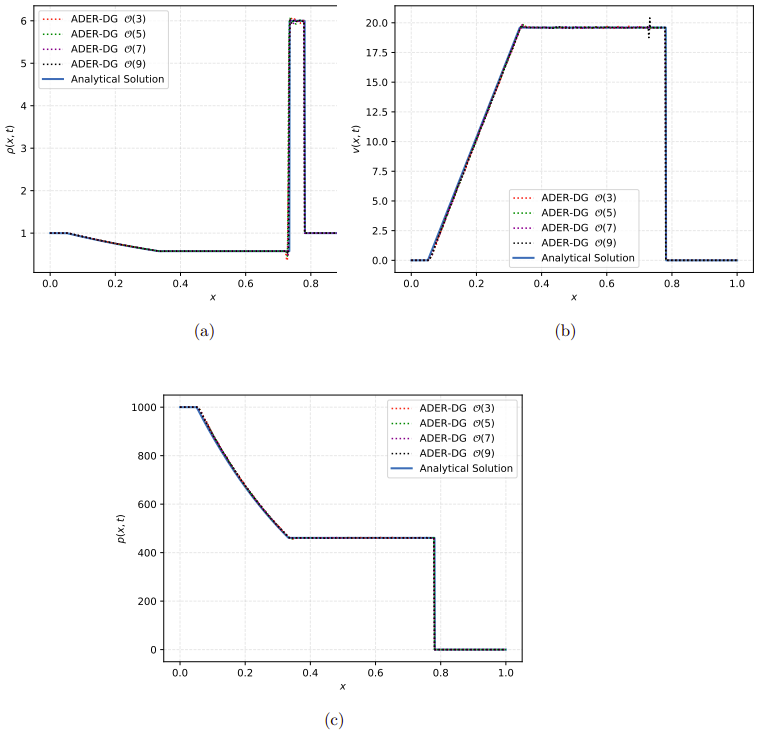}
  \caption{Sod shock tube profiles at $t = 0.012$ for density,
           velocity, and pressure. ADER-DG results at orders
           3, 5, 7, and 9 are compared against the exact
           Riemann solution.}
  \label{fig:sod_profiles}
\end{figure}

The profiles in Figure~\ref{fig:sod_profiles} compare density,
velocity, and pressure for all four orders against the exact
Riemann solution at $t = 0.012$. All orders place the
rarefaction, contact, and shock in the expected sequence. The
rarefaction fan between $x \approx 0.05$ and $x \approx 0.35$
becomes sharper as the order increases, while shock and contact
positions remain accurate across the set.

The same comparison exposes the local trade-off near
discontinuities. Around the contact at $x \approx 0.73$, the
seventh-order solution is sharp and nearly free of diffusion;
the ninth-order solution, by contrast, develops localised
dispersive ripples. These Gibbs-type oscillations persist
because the candidate DG polynomial remains physically
admissible. The limiter intervenes only when admissibility is
threatened, replacing the affected cell with a monotone
finite-volume update rather than filtering every oscillation.

In smooth regions, the density error decreases consistently
with polynomial order. The convergence behaviour is in
agreement with the theoretical rate of $N + 1$ for \aderdg{}
schemes \citep{Dumbser2008}. A systematic convergence study
with $L^1$ and $L^2$ errors across multiple mesh resolutions
will be added in a later preprint update. Here, the Sod test is
used more narrowly: it checks wave-pattern fidelity and the
behaviour of the limiter under an extreme pressure jump.

The extreme pressure ratio of $10^5$ in this configuration
mimics, in a controlled one-dimensional setting, the strong
compressive fronts generated during the free-expansion phase
of supernova remnants.

\subsection{Shu--Osher Problem}
\label{sec:shu_osher}

The Shu--Osher problem extends the Riemann setting by placing
a sinusoidal density perturbation ahead of a moving shock
\citep{ShuOsher1989}. The shock propagates into this modulated
medium and generates a complex post-shock oscillatory pattern.
This exercises spatial resolution in a way the Sod test
cannot: both a sharp front and fine-scale smooth structures
must be resolved simultaneously. The initial conditions
are in Table~\ref{tab:shu_ic}.

\begin{table}[htbp]
  \centering
  \caption{Initial conditions for the Shu--Osher problem.
           Domain $[-5,5]$, discontinuity at $x = -4.5$,
           final time $t = 5$.}
  \label{tab:shu_ic}
  \begin{tabular}{lccc}
    \toprule
    State & $\rho$ & $u$ & $p$ \\
    \midrule
    Left  & $1.515695$ & $0.523346$ & $1.805$ \\
    Right & $1 + 0.1\sin(20\pi x)$ & $0.0$ & $1.0$ \\
    \bottomrule
  \end{tabular}
\end{table}

We run orders $N = 2, 4, 6$ with AMR up to one refinement
level. No exact analytical solution exists, so we use a
seventh-order simulation on a 5000-cell uniform grid as
the reference.

\begin{figure}[H]
  \centering
  \includegraphics[width=0.92\textwidth]{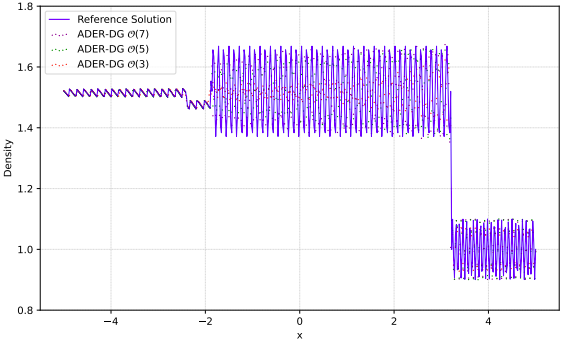}
  \caption{Shu--Osher density profiles at $t = 5$ over the full
           domain. ADER-DG results at orders 3, 5, and 7 are
           compared against a high-resolution reference solution.}
  \label{fig:shu_profiles}
\end{figure}

The density profiles in Figure~\ref{fig:shu_profiles} make the
order dependence clear at $t = 5$. Third order damps the
post-shock oscillations. Fifth order recovers their general
shape but still underestimates peak amplitudes. Seventh order
follows the reference most closely throughout the oscillatory
region, although a small phase shift remains visible at all
orders.
It accumulates because even minor differences in numerical
shock speed displace the whole post-shock pattern over five
time units. The shift does not reflect inaccuracy in
amplitude or frequency resolution.

The detailed post-shock structure points to the same conclusion:
seventh order captures the amplitude and spacing of the
oscillations, but the whole pattern sits slightly ahead of the
reference. For this comparison, a phase-insensitive diagnostic is more
informative than a pointwise norm.

\subsubsection*{Phase-insensitive error analysis via
                Shannon entropy}

Standard $L^1$ and $L^2$ norms are not suitable here.
A small phase shift produces large norm values even when
amplitude and frequency are accurately captured. We compare the
solutions through their amplitude distributions rather than their
pointwise values.

For the entropy analysis we restrict the comparison to the
post-shock oscillatory region, excluding the undisturbed
upstream state and the shock jump itself. The restricted
interval is where phase errors are most visible, but the main
question is whether the numerical method reproduces the
resolved amplitude distribution. For each
scheme we build a normalised histogram of density values over
the same spatial interval and using the same fixed binning as
the reference solution. The resulting probabilities define an
empirical amplitude distribution rather than a pointwise error
measure. Tests with nearby bin counts give the same ordering
of schemes, so the interpretation depends on the relative
broadening and entropy of the distributions, not on a
particular choice of bin width.

We compute the amplitude histogram of the post-shock density
field for each order and for the reference. The histogram
quantifies how well each scheme reproduces the statistical
spread of the oscillatory signal, independent of phase.
We then measure the Shannon entropy of each distribution
\citep{Shannon1948},
\begin{equation}
  H = -\sum_{k} p_k \log_2 p_k,
  \label{eq:shannon}
\end{equation}
where $p_k$ is the normalised frequency of the $k$-th bin.
Higher entropy corresponds to a broader and more structured
amplitude distribution, hence to a larger fraction of the
shock--entropy imprint being resolved. Because only the
amplitude distribution enters, spatial phase is discarded by
construction.

\begin{figure}[H]
  \centering
  \begin{subfigure}{0.55\textwidth}
    \centering
    \includegraphics[width=\textwidth]{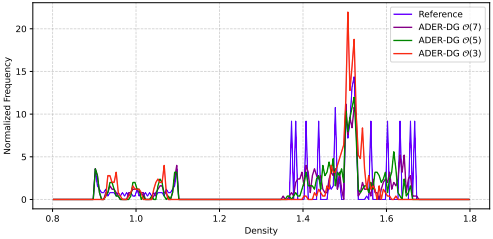}
    \caption{Normalised amplitude histograms.}
    \label{fig:shu_entropy_histograms}
  \end{subfigure}
  \hfill
  \begin{subfigure}{0.40\textwidth}
    \centering
    \includegraphics[width=\textwidth]{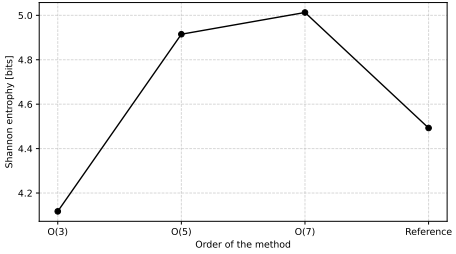}
    \caption{Entropy versus order.}
    \label{fig:shu_entropy_order}
  \end{subfigure}
  \caption{Phase-insensitive Shu--Osher error analysis based on
           density-amplitude distributions. The histogram panel
           compares numerical orders against the reference, while
           the entropy panel summarises the recovered structural
           diversity as Shannon entropy.}
  \label{fig:shu_entropy}
\end{figure}

Figure~\ref{fig:shu_entropy} summarises this comparison. The
third-order histogram is narrow and modal, a signature of
excessive diffusion. As the order increases, the histograms
broaden toward the reference distribution. The corresponding
Shannon entropies, listed in Table~\ref{tab:shu_entropy}, rise
monotonically with order. Most of the gain occurs between
orders 3 and 5; the smaller increment from orders 5 to 7
suggests diminishing returns once dispersive effects begin to
compete with additional resolution.
The reference entropy is the upper bound a perfect numerical
solution would reach at this resolution.

\begin{table}[htbp]
  \centering
  \caption{Shannon entropy $H$ of the post-shock density
           amplitude distribution for the Shu--Osher problem.
           Higher values indicate greater structural fidelity.}
  \label{tab:shu_entropy}
  \begin{tabular}{lc}
    \toprule
    Scheme & $H$ (bits) \\
    \midrule
    ADER-DG $N=2$ (order 3) & $4.21$ \\
    ADER-DG $N=4$ (order 5) & $4.67$ \\
    ADER-DG $N=6$ (order 7) & $4.84$ \\
    Reference               & $4.98$ \\
    \bottomrule
  \end{tabular}
\end{table}

In this setting, Shannon entropy acts as a phase-insensitive
quality metric for \aderdg{} schemes in the shock--entropy
interaction regime. It complements norm-based comparisons when
phase drift is part of the numerical response rather than a
simple loss of amplitude accuracy.

The Shu--Osher configuration models the interaction of
supernova-driven shocks with pre-existing density
inhomogeneities in the interstellar medium, where entropy
perturbations modulate the post-shock flow structure.

\subsection{Woodward--Colella Blast Wave}
\label{sec:woodward_colella}

The Woodward--Colella problem is the most demanding of the
three \citep{WoodwardColella1984}. Two strong shocks propagate
inward from high-pressure boundaries, collide near the
centre, reflect, and produce a transient multi-wave structure
with interacting shocks, contacts, and rarefactions. The
pressure ratio is $10^5$. The initial conditions are in
Table~\ref{tab:wc_ic}.

\begin{table}[htbp]
  \centering
  \caption{Initial conditions for the Woodward--Colella
           blast wave. Domain $[0,1]$, final time $t=0.038$.
           Density $\rho = 1$ and velocity $u = 0$ everywhere.}
  \label{tab:wc_ic}
  \begin{tabular}{lc}
    \toprule
    Region & $p$ \\
    \midrule
    $x < 0.1$             & $1000.0$ \\
    $0.1 \leq x \leq 0.9$ & $0.01$   \\
    $x > 0.9$             & $100.0$  \\
    \bottomrule
  \end{tabular}
\end{table}

No analytical solution exists. We construct a reference
solution with a seventh-order \aderdg{} scheme at
$\Delta x = 10^{-5}$. That simulation required more than
ninety hours of wall-clock time on a single node. The cost
illustrates why matching such a reference by brute-force
refinement is unattractive, especially for lower-order schemes.
Production simulations use $\Delta x = 10^{-3}$ with two
AMR levels.

\begin{figure}[!htbp]
  \centering
  \includegraphics[width=0.80\textwidth]{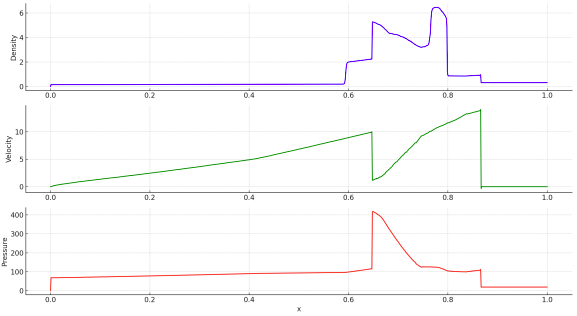}
  \caption{High-resolution reference solution for the
           Woodward--Colella blast wave at $t = 0.038$,
           showing density, velocity, and pressure together with
           the principal wave structures.}
  \label{fig:wc_reference}
\end{figure}

The reference solution at $t = 0.038$ is shown in
Figure~\ref{fig:wc_reference}. The dominant
features are two reflected shock fronts near $x \approx 0.647$
and $x \approx 0.867$, contact discontinuities near
$x \approx 0.763$--$0.800$, and outward rarefaction fans
near both boundaries. These structures are the benchmark
for the production simulations.

\begin{figure}[H]
  \centering
  \begin{subfigure}{0.32\textwidth}
    \centering
    \includegraphics[width=\textwidth]{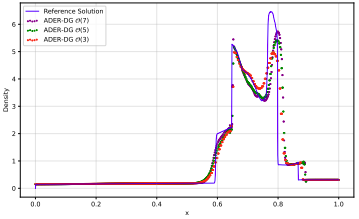}
    \caption{Density.}
    \label{fig:wc_density}
  \end{subfigure}
  \hfill
  \begin{subfigure}{0.32\textwidth}
    \centering
    \includegraphics[width=\textwidth]{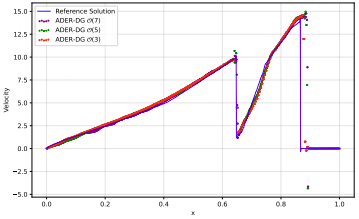}
    \caption{Velocity.}
    \label{fig:wc_velocity}
  \end{subfigure}
  \hfill
  \begin{subfigure}{0.32\textwidth}
    \centering
    \includegraphics[width=\textwidth]{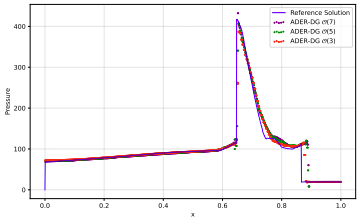}
    \caption{Pressure.}
    \label{fig:wc_pressure}
  \end{subfigure}
  \caption{Woodward--Colella blast wave profiles at $t = 0.038$.
           ADER-DG results at orders 3, 5, and 7 are compared
           against the high-resolution reference solution for
           density, velocity, and pressure.}
  \label{fig:wc_profiles}
\end{figure}

The production runs in Figure~\ref{fig:wc_profiles} compare
orders $N = 2, 4, 6$. Third order captures the global wave
layout but smears shocks and contact surfaces. Fifth order
places the primary shocks accurately and resolves most contacts.
Seventh order follows the reference across density, velocity,
and pressure, including the secondary contacts and the transient
central region.
The subcell limiter is active across a large fraction of
the domain throughout the interaction phase. This behaviour is
consistent with the problem setup: extreme pressure gradients
trigger the admissibility check repeatedly, and the limiter
responds locally without degrading the high-order solution
elsewhere.

The blast wave dynamics here mirror, in one spatial
dimension, the shock focusing and vorticity deposition that
occur in the interior evolution of supernova remnants and
stellar wind-blown cavities.

\begin{table}[H]
  \centering
  \footnotesize
  \caption{Summary of the one-dimensional benchmark results.}
  \label{tab:results_1d_summary}
  \begin{tabularx}{\textwidth}{p{0.16\textwidth} p{0.24\textwidth} X}
    \toprule
    Benchmark & Numerical challenge & Main finding \\
    \midrule
    Sod &
    Rarefaction, contact, and shock at pressure ratio $10^5$ &
    All orders recover the correct wave pattern; higher order reduces
    smooth-region diffusion, while the highest order exposes local
    dispersive oscillations near discontinuities. \\
    \addlinespace
    Shu--Osher &
    Shock interaction with entropy waves and phase-sensitive oscillations &
    Seventh order best reproduces the oscillatory amplitude structure;
    Shannon entropy provides a phase-insensitive measure of recovered
    fine-scale content. \\
    \addlinespace
    Woodward--Colella &
    Strong shock interactions, reflections, contacts, and limiter activation &
    Seventh order remains closest to the high-resolution reference while
    preserving robustness in the most compressive regime tested here. \\
    \bottomrule
  \end{tabularx}
\end{table}

Taken together, the one-dimensional tests isolate the regime
that motivates this validation: shocks and contacts coexist with
entropy structure, large pressure jumps, and repeated limiter
activation. In that regime, high-order \aderdg{} with subcell
limiting reproduces the standard Riemann wave pattern, retains
post-shock oscillatory content, and remains stable in the
Woodward--Colella blast wave. These results make the method a
credible candidate for astrophysical flows in which the relevant
structure is not always captured by pointwise errors alone.

\FloatBarrier

\section{Two-Dimensional Benchmark Tests}
\label{sec:results_2d}

The one-dimensional tests in Section~\ref{sec:results_1d}
exercise wave interactions along a single axis. In two space
dimensions, additional effects enter: waves can propagate
obliquely, interfaces can interact geometrically, and
hydrodynamic instabilities can grow from genuinely
multidimensional gradients. The benchmarks in this section target
these effects in two complementary limits. The first contains only
contact discontinuities and tangential shear. The second adds a
strong oblique shock that strikes a density interface and drives
multiscale instability growth. Together, the tests extend the
validation beyond colinear wave dynamics. Both calculations use
polynomial order $N = 4$ --- a representative fifth-order
configuration --- with AMR and the subcell limiter active
throughout.

\subsection{Contact-Driven Vortex Sheet Dynamics}
\label{sec:vortex_sheet}

The first problem is designed to isolate contact discontinuities
and tangential shear, with no shock forcing. The domain $[0,1]^2$
is divided into four quadrants at $x = 0.5$ and $y = 0.5$, each
with its own velocity state. Density and pressure are constant
inside each quadrant. The density jumps define the contact
surfaces, while tangential velocity jumps across those contacts
provide the initial shear. Table~\ref{tab:vortex_ic} gives the
corresponding states.

\begin{table}[htbp]
  \centering
  \caption{Initial conditions for the vortex sheet test.
           Domain $[0,1]^2$, interfaces at $x = y = 0.5$.
           Pressure $p = p_0$ is uniform. Four values of
           $p_0$ are studied: $1.0,\, 0.75,\, 0.5,\, 0.25$.}
  \label{tab:vortex_ic}
  \begin{tabular}{lcccc}
    \toprule
    Quadrant & $\rho$ & $u$ & $v$ & $p$ \\
    \midrule
    Top-left $(x < 0.5,\; y > 0.5)$      & $1$ & $\phantom{-}0.75$ & $-0.5$              & $p_0$ \\
    Top-right $(x > 0.5,\; y > 0.5)$     & $2$ & $\phantom{-}0.75$ & $\phantom{-}0.5$  & $p_0$ \\
    Bottom-right $(x > 0.5,\; y < 0.5)$  & $1$ & $-0.75$           & $\phantom{-}0.5$  & $p_0$ \\
    Bottom-left $(x < 0.5,\; y < 0.5)$   & $3$ & $-0.75$           & $-0.5$              & $p_0$ \\
    \bottomrule
  \end{tabular}
\end{table}

The shear dynamics are described by the two-dimensional
vorticity field,
\begin{equation}
  \omega = \frac{\partial v}{\partial x}
         - \frac{\partial u}{\partial y},
  \label{eq:vorticity}
\end{equation}
and the baroclinic contribution to its evolution is
\begin{equation}
  \frac{D\omega}{Dt}
  = \frac{1}{\rho^2}
    \left(\nabla\rho \times \nabla p\right)_z.
  \label{eq:baroclinic}
\end{equation}
The right-hand side of~\eqref{eq:baroclinic} is the baroclinic
torque. In this setup the initial pressure is spatially uniform,
so $\nabla p = 0$ at $t = 0$ and there is no initial baroclinic
source. Vorticity is instead introduced by the imposed tangential
jumps at the quadrant interfaces. The subsequent motion is then a
nonlinear reorganisation of these vortex sheets rather than the
result of impulsive baroclinic deposition.

Four simulations are run with
$p_0 \in \{1.0,\, 0.75,\, 0.5,\, 0.25\}$, while the resolution is
kept fixed at $\Delta x = 10^{-3}$ with two AMR levels. Decreasing
$p_0$ raises the Mach number of the tangential flow and moves the
problem progressively farther from the nearly incompressible
limit.

\begin{figure}[!htbp]
  \centering
  \includegraphics[width=0.80\textwidth]{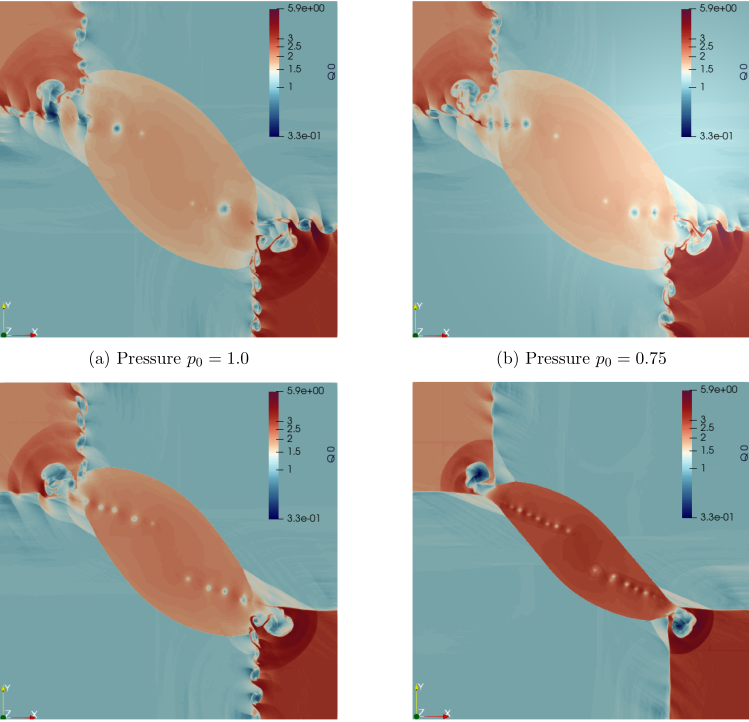}
  \caption{Density fields for the contact-driven vortex sheet
           test at four background pressures,
           $p_0 = 1.0,\,0.75,\,0.5,\,0.25$. Lower pressure
           increases compressibility and strengthens the density
           filamentation associated with vortex-sheet roll-up.}
  \label{fig:vortex_density}
\end{figure}

Figure~\ref{fig:vortex_density} compares the four pressure values
at the same final time. In each case, the oblique shear layers
wind into alternating high- and low-density streaks as the flow is
rotated and stretched. The value of $p_0$ controls both the
strength and the internal structure of this pattern. For
$p_0 = 1.0$, the motion remains close to incompressible and the
vortical structures are compact, with only mild deformation. As
$p_0$ is reduced, density contrasts sharpen and secondary vortices
emerge near the outer corners of the domain. Their shape is
consistent with Kelvin--Helmholtz-type growth, produced as the
main vortical structures interact with outward-propagating
fluctuations. The lowest-pressure case, $p_0 = 0.25$, produces the
most developed vortex system, with the largest number of
secondary structures and the strongest density filamentation.

The numerical solution follows this progression in all four
cases. Smooth interior regions remain free of visible grid-scale
noise, and the contact surfaces stay sharp throughout the
evolution. Because no shocks are present, the problem gives a
direct check on whether the limiting \aderdg{} formulation can preserve
contact layers without adding enough diffusion to erase the rolled-up
vortex structure.

The same type of shear-driven roll-up appears in several
astrophysical settings. Tangential velocity discontinuities at jet
boundaries, and in shear-dominated regions of the interstellar
medium, can undergo similar Kelvin--Helmholtz growth, producing
filamentary structure and turbulent mixing.

\subsection{Shock--Interface Interaction}
\label{sec:shock_interface}

The second problem adds the compressive forcing that is absent
from the vortex-sheet test. A strong shock propagates obliquely
into a stratified density field, producing a transmitted shock, a
reflected wave, and a deforming material interface. In this
configuration, unlike the previous one, the baroclinic
term~\eqref{eq:baroclinic} is active. Across the interface,
$\nabla\rho$ and $\nabla p$ are strongly misaligned, so the
baroclinic torque deposits vorticity directly on the contact
surface. This is the mechanism underlying Richtmyer--Meshkov
growth \citep{Richtmyer1960, Meshkov1969}: impulsive vorticity
deposition drives the interface toward a bubble--spike morphology,
which later feeds Kelvin--Helmholtz-like roll-up at the tips.
Table~\ref{tab:shock_ic} lists the initial states.

\begin{figure}[t]
  \centering
  \includegraphics[width=0.80\textwidth]{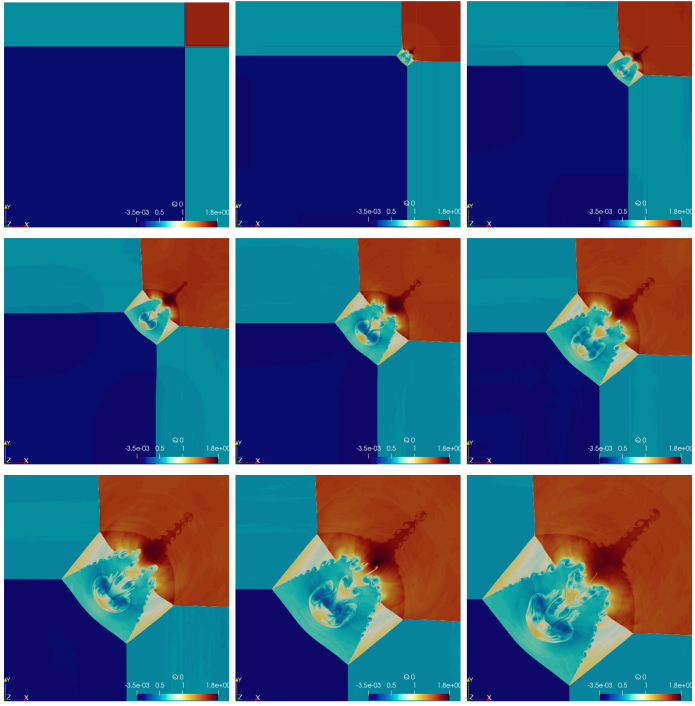}
  \caption{Temporal evolution of the density field in the
           shock--interface interaction. The snapshots show shock
           propagation, interface deformation, Richtmyer--Meshkov
           growth, and secondary Kelvin--Helmholtz roll-up.}
  \label{fig:shock_interface}
\end{figure}

\begin{table}[htbp]
  \centering
  \caption{Initial conditions for the shock--interface
           interaction. Domain $[0,1]^2$, interfaces at
           $x = y = 0.8$. Non-reflecting outflow conditions
           on all boundaries. Resolution $\Delta x =
           \Delta y = 10^{-3}$.}
  \label{tab:shock_ic}
  \begin{tabular}{lcccc}
    \toprule
    Region & $\rho$ & $u$ & $v$ & $p$ \\
    \midrule
    1 $(x > 0.8,\; y > 0.8)$ & $1.500$  & $0.000$ & $0.000$ & $1.500$ \\
    2 $(x < 0.8,\; y > 0.8)$ & $0.5323$ & $1.206$ & $0.000$ & $0.300$ \\
    3 $(x < 0.8,\; y < 0.8)$ & $0.138$  & $1.206$ & $1.206$ & $0.029$ \\
    4 $(x > 0.8,\; y < 0.8)$ & $0.5323$ & $0.000$ & $1.206$ & $0.300$ \\
    \bottomrule
  \end{tabular}
\end{table}
The snapshots in Figure~\ref{fig:shock_interface} show the
temporal evolution of the density field. The initial shock front
moves toward the lower-left and strikes the contact interfaces
obliquely. After impact, the wave separates into transmitted and
reflected components, while the material interface deforms and
starts to roll up. At later times the interface develops the
expected Richtmyer--Meshkov morphology: heavy-fluid spikes extend
into the lighter region, and lighter-fluid bubbles rise into the
heavier one. Kelvin--Helmholtz-like roll-up then appears at the
spike tips as secondary shear between the two fluids amplifies.
As these features grow and interact, the density field becomes
multiscale, combining large wave structures with smaller vortical
patterns.

The scheme captures this evolution without introducing spurious
oscillations in the smooth regions between wave structures. The
subcell limiter is triggered at shock fronts and along the
deforming interface, where steep gradients form, but the
high-order update is retained in smooth downstream regions.
Entropy waves, visible as sharp density layers advected by the
flow, remain well resolved across the domain. This behaviour
reflects the intended division of labour in the \textit{a
posteriori} limiter: troubled zones receive a robust finite-volume
update, while smooth regions continue to use the high-order DG
evolution.

The same physical mechanism --- baroclinic vorticity deposition at
a shock-accelerated density interface --- is relevant when
supernova blast waves encounter dense molecular clouds in the
interstellar medium. In that setting, RMI-driven mixing can help
shape filamentary structures.

The one-dimensional tests of Section~\ref{sec:results_1d} showed
that the scheme resolves the three Euler wave families along a
single axis. The present tests extend that assessment to
multidimensional interface dynamics. The vortex-sheet problem
probes contact surfaces and shear-driven roll-up without shock
forcing; the shock--interface problem adds impulsive vorticity
deposition, transmitted shocks, and secondary
Kelvin--Helmholtz-like structure. In both cases, the \aderdg{}
scheme with subcell limiting and AMR handles smooth and non-smooth
features within the same calculation. Related interface geometries
also occur outside astrophysics, for example in atmospheric
density fronts and gravity currents. The analogy is geometric and
dynamical rather than one of scale, and is developed further in
Section~\ref{sec:discussion}.

\FloatBarrier

\section{Discussion: Numerical Findings and Physical Relevance}
\label{sec:discussion}

The benchmark sequence in Sections~\ref{sec:results_1d}
and~\ref{sec:results_2d} was designed to reduce the ambiguity of the
assessment one step at a time. It tests the \aderdg{} implementation in
flows where smooth waves, shocks, contacts, and multidimensional
interfaces appear in different combinations. The useful point is not just
that each test can be run, but that the same numerical ingredients behave
consistently as the character of the flow changes. For that reason, the
physical interpretation is kept narrow: the results concern the inviscid,
non-relativistic Euler limit, and the analogies below should be read
within that limit.

\subsection{Numerical Findings in Context}
\label{sec:findings}

The five benchmark problems are intended as an escalation in physical and
numerical difficulty, rather than as independent demonstrations. Sod checks
the basic ordering and propagation of the Euler wave families. Shu--Osher
adds a shock moving through smooth entropy structure. Woodward--Colella
then asks whether the method remains robust during repeated strong-shock
interactions. The vortex sheet removes shock forcing and isolates contact
preservation under tangential shear. Finally, the shock--interface problem
combines shocks, density interfaces, baroclinic vorticity deposition, and
secondary instability growth.

Taken together, the tests support three observations. First, in one
dimension the solver recovers the expected Euler wave structure:
rarefactions, contacts, and shocks appear in the correct sequence, and
higher polynomial order improves smooth or oscillatory regions in the way
expected for high-order \aderdg{} schemes \citep{Dumbser2008}. Second, in
strongly compressive cases, the \textit{a posteriori} subcell limiter acts
as a reliable fallback without noticeably washing out the smooth parts of
the solution. Third, in two dimensions, the method keeps interfaces sharp
enough to resolve vortical structures generated by shear and by
baroclinic shock--interface interaction.

Table~\ref{tab:synthesis} summarises the role of each benchmark in
this validation argument.

\begin{table}[htbp]
  \centering
  \footnotesize
  \caption{Role of each benchmark in the validation argument.}
  \label{tab:synthesis}
  \begin{tabularx}{\textwidth}{p{0.20\textwidth} X p{0.25\textwidth}}
    \toprule
    Benchmark & Role in the validation & Physical analogue \\
    \midrule
    Sod &
    Tests the ordering and propagation of rarefactions, contacts,
    and shocks under a strong pressure jump. &
    Idealised blast-wave front \\
    \addlinespace
    Shu--Osher &
    Tests the ability of higher order to preserve post-shock
    oscillatory structure in an inhomogeneous flow. &
    Shocks in inhomogeneous gas \\
    \addlinespace
    Woodward--Colella &
    Tests robustness under repeated strong shock interaction,
    reflection, and limiting. &
    Confined blast-wave dynamics \\
    \addlinespace
    Vortex sheet &
    Tests contact preservation and shear-driven roll-up without
    shock forcing. &
    Jet shear layers and KHI \\
    \addlinespace
    Shock--interface &
    Tests baroclinic vorticity deposition and shock-driven
    interface growth in two dimensions. &
    Shock--cloud interaction \\
    \bottomrule
  \end{tabularx}
\end{table}

Read together, the entries in Table~\ref{tab:synthesis} make the main
argument of the validation. The one-dimensional tests check wave-pattern
fidelity and robustness under strong compression; the two-dimensional
tests show that the same ingredients remain useful when geometry, shear,
and baroclinic vorticity enter. No single benchmark would be enough to
support that conclusion on its own.

\subsection{Astrophysical Relevance}
\label{sec:astro_relevance}

The clearest physical connection is with idealised astrophysical
hydrodynamics. During the early, non-radiative evolution of
supernova remnants, the Euler equations provide the standard
leading-order description of freely expanding ejecta and of the
subsequent adiabatic Sedov--Taylor stage
\citep{Chevalier1982, Truelove1999}. The Sod and
Woodward--Colella tests do not model a remnant directly, but they
probe the same inviscid ingredients: strong pressure contrasts,
blast-wave propagation, reflected shocks, contacts, and
rarefactions. Shu--Osher adds the interaction of a shock with a
pre-existing density modulation, a mechanism connected to
post-shock structure generation in inhomogeneous astrophysical
media and core-collapse supernova flows \citep{Abdikamalov2018}.

The shock--interface test is particularly relevant to blast waves
encountering dense clouds in the interstellar medium. The present
configuration is idealised: it excludes stratification, magnetic
fields, radiative cooling, and realistic cloud geometry. Even so,
it isolates the same leading-order inviscid process that drives
Richtmyer--Meshkov growth in those systems: impulsive baroclinic
vorticity deposition at a shock-accelerated density interface.

Non-relativistic astrophysical jets provide a second natural
connection. A collimated jet propagating through an ambient medium
contains both boundary shear layers and a shocked head region
\citep{Hardee1979, NormanWinkler1985}. The vortex-sheet benchmark
isolates the first geometry, where tangential velocity jumps drive
Kelvin--Helmholtz-type roll-up. The shock--interface benchmark
isolates the second, where an oblique shock deforms a density
interface and deposits vorticity. The results support the implementation as a starting point for idealised
inviscid studies of non-relativistic jet-boundary dynamics, while stopping
well short of a complete jet model.

\subsection{Broader Fluid-Dynamical Analogues}
\label{sec:broader}

These mechanisms are not uniquely astrophysical. Atmospheric
density currents can be idealised locally as sharp interfaces
between air masses of different density, and Kelvin--Helmholtz
instability is one recognised route to clear-air turbulence in
stratified shear layers \citep{Klemp1994, Miles1961, Fritts2003}.
These examples are included only to mark the broader mathematical
setting of the tests. They are not additional validation cases:
rotation, stratification, moisture, and external forcing are not
included in the present Euler benchmarks.

\subsection{Limitations}
\label{sec:limitations}

The validation remains intentionally limited. All benchmarks are
one- or two-dimensional, so genuinely three-dimensional effects such
as vortex stretching, turbulent cascades, and azimuthal instability
modes are outside the present evidence. The governing equations are
the ideal inviscid Euler equations; viscosity, magnetic fields,
radiative cooling, gravity, and relativistic corrections are not
included. The astrophysical and atmospheric connections above
therefore apply only to idealised, inviscid, non-relativistic
limits.

The assessment is also internal to this implementation. A direct
comparison with established astrophysical solvers such as PLUTO
\citep{Mignone2007} or FLASH \citep{Fryxell2000} on the same test
suite would be needed to make a competitive performance claim. In
addition, this version does not report a formal multi-resolution
$L^1$ or $L^2$ convergence table, nor does it quantify the fraction
of cells treated by the subcell limiter. A future quantitative study
should add convergence rates and troubled-cell statistics. Within these limits, the benchmark sequence supports the solver as a
robust high-order Euler implementation for idealised shock- and
interface-dominated flows.

\section{Conclusions}
\label{sec:conclusions}

We implemented and tested a high-order \aderdg{} solver for the
compressible Euler equations in \exahype{}. The solver combines a
high-order \aderdg{} polynomial representation, a local space--time DG
predictor, \textit{a posteriori} subcell finite-volume limiting, and
adaptive mesh refinement. The codes and datasets are publicly available, which should
make the benchmark sequence straightforward to reproduce or extend.

The tests were chosen to build up the difficulty step by step. In one
dimension they check the ordering of Euler wave families, the interaction
of a shock with entropy structure, and the behaviour of the limiter under
repeated strong-shock reflection. In two dimensions they add contact
preservation, shear-driven roll-up, baroclinic vorticity deposition, and
Richtmyer--Meshkov-type interface growth. Across this sequence, the
solver behaves as intended: high-order resolution is retained in smooth
parts of the flow, while the subcell limiter provides a stable fallback
near shocks and steep gradients.

The most direct physical use of these tests is in idealised astrophysical
hydrodynamics. Blast fronts, shock--entropy interactions, shear layers,
and shock-accelerated interfaces all appear in simplified descriptions of
early supernova-remnant evolution, shock--cloud interaction, and
non-relativistic jet-boundary dynamics. The present results should be
read in that restricted sense: they support the inviscid,
non-relativistic Euler limit, not a complete model including cooling,
magnetic fields, gravity, or relativistic effects.

A natural next step is a more quantitative comparison using formal
multi-resolution $L^1$ and $L^2$ errors, together with statistics on
troubled-cell activation. Direct comparisons with established
astrophysical hydrodynamics codes on the same tests would also clarify
where this implementation is most competitive. Within the scope tested
here, the benchmark suite supports the solver as a robust high-order tool
for idealised flows dominated by shocks, contacts, and multidimensional
interfaces.

\section*{Declaration of competing interest}
The authors declare that they have no known competing financial interests
or personal relationships that could have appeared to influence the work
reported in this paper.

\section*{Data availability}
All simulation codes and datasets produced in this study are publicly
available. Repository links are provided in \cref{sec:appendix}.

\begin{appendices}
\section{Code and Data Repository}
\label{sec:appendix}

The codes and datasets used for the benchmark calculations are
publicly available through the resources listed in
Table~\ref{tab:code_list}. They include analytical reference tools,
predictor notebooks, and the \exahype{} configurations used for the
one- and two-dimensional Euler benchmarks discussed in
Sections~\ref{sec:results_1d} and~\ref{sec:results_2d}.

\begin{table}[H]
  \centering
  \footnotesize
  \caption{Public code and data resources used in this study.}
  \label{tab:code_list}
  \begin{tabularx}{\textwidth}{p{0.36\textwidth} p{0.16\textwidth} X}
    \toprule
    Resource & Language / framework & Public link \\
    \midrule
    Analytical solution to the Sod shock tube problem &
    Python &
    \url{https://github.com/anmsuarezma/Master_Thesis/blob/main/riemannSolution.ipynb} \\
    \addlinespace
    Galerkin predictor of order 3 for the linear advection equation &
    Python &
    \url{https://github.com/anmsuarezma/Master_Thesis/blob/main/galerkin3Order.ipynb} \\
    \addlinespace
    Lagrange nodal points for interpolation in DG schemes &
    Python &
    \url{https://github.com/anmsuarezma/Master_Thesis/blob/main/polynomialLagrangeNodalPoints.ipynb} \\
    \addlinespace
    ADER-DG Sod-type benchmark configuration using \exahype{} &
    \exahype{} &
    \url{https://github.com/anmsuarezma/Master_Thesis/blob/main/sod2D\%20.tar.xz} \\
    \addlinespace
    Numerical solution to the Shu--Osher problem &
    \exahype{} &
    \url{https://osf.io/5kamd} \\
    \addlinespace
    Woodward--Colella blast-wave simulation &
    \exahype{} &
    \url{https://osf.io/hbfc3} \\
    \addlinespace
    Vortex-sheet dynamics with adaptive refinement &
    \exahype{} &
    \url{https://osf.io/kdjzb} \\
    \addlinespace
    Shock--interface interaction in compressible hydrodynamics &
    \exahype{} &
    \url{https://osf.io/2v9xd} \\
    \bottomrule
  \end{tabularx}
\end{table}

\end{appendices}

\clearpage

\printbibliography

\end{document}